# On-the fly AES Decryption/Encryption for Cloud SQL Databases

Position Statement

Sushil Jajodia[1], Witold Litwin[2], Thomas Schwarz[3]

[1] George Mason University, Fairfax, Virginia, USA
jajodia@gmu.edu
[2] Université Paris Dauphine, Paris, France[*]
witold.litwin@dauphine.fr
[3] Universidad Centroamericana, San Salvador, El Salvador
tschwarz@uca.edu.sv

June 15, 2015[a]

**Abstract.** We propose the client-side AES256 encryption for a cloud SQL DB. A column ciphertext is deterministic or probabilistic. We trust the cloud DBMS for security of its run-time values, e.g., through a moving target defense. The client may send AES key(s) with the query. These serve the on-the-fly decryption of selected ciphertext into plaintext for query evaluation. The DBMS clears the key(s) and the plaintext at the query end at latest. It may deliver ciphertext to decryption enabled clients or plaintext otherwise, e.g., to browsers/navigators. The scheme functionally offers to a cloud DBMS capabilities of a plaintext SQL DBMS. AES processing overhead appears negligible for a modern CPU, e.g., a popular Intel I5. The deterministic encryption may have no storage overhead. The probabilistic one doubles the DB storage. The scheme seems the first generally practical for an outsourced encrypted SQL DB. An implementation sufficient to practice with appears easy. An existing cloud SQL DBMS with UDF support should do.

Motto: (1) *If the door doesn't open, try a window.* Old days saying. (2) *In Cloud we trust.* Nowadays saying.

## 1. Introduction

A cloud outsourced database (DB) evidently welcomes the client-side encryption. The traditional security paradigm for a cloud DBs is that any cloud storage is simply (somehow uniformly) insecure against unwelcome reading. Sensitive client data should never ever be at the cloud. In [JLS5] we proposed a new paradigm. It trusts the cloud DBMS software secure for its run-time (fugitive, furtive…) values (variables) with sensitive client data. These are believed protected even against disclosure by insiders of the cloud administration. Only the (persistent) cloud files, are eventually insecure.

Our paradigm seemed new for a cloud DBMS. It comes nevertheless hardly out of the blue. Billions of users similarly trust their favorite industrial OSs and browsers. We all have no choice but to readily type-in various sensitive data, e.g., all passwords, credit card numbers, you name it. Trusting firmly the values provided secure in some run-time variables only. Wiped out immediately after the use. We are many, in contrast, to politely refuse the Google Chrome offering to turn these persistent. Likewise, we often do not trust the security of (persistent) plaintext files. Usually we provide the sensitive ones with the password encryption at least, [TCG]. Our paradigm simply transposes this overwhelming reality to the cloud DB arena.

Besides, the only difference between both paradigms is that the new one trusts the DBMS at the cloud and the one at the client, while the traditional one focuses on the latter only. In practice, both should be third-part made, by industrial players in general. Both are also obviously Internet accessible. No reason then for trusting more one installation than the other. E.g., there is no reason to trust SQL

---

[*] Work performed while visiting GMU Center for Secure Information Systems (CSIS), June - July 2015.
[a] Updated on December 20, 2015

Server secure for run-time values locally and *a priori* distrust it elsewhere. We all know also that attacks are mostly remote now anyhow. Life shows finally that the insider attacks are *a priori* about as likely at a client node as at the cloud.

To be generally practical, the cloud SQL DBMS should evaluate most queries at the cloud. This especially concerns the select-project-join (SPJ) queries, usual for the online transaction processing (OLTP). If the cloud processing is not feasible, an impractical transfer to the client may be the only possibility. An SQL query may also have value expressions, These are especially frequent for the popular Online Analytical Processing (OLAP), where aggregations prevail. Under the traditional paradigm, to avoid the impractical transfers to the client, a fully homomorphic encryption scheme is functionally necessary. Such a scheme should be also sufficiently fast. Ideally, encrypted data (ciphertext) processing should be about as fast as the plaintext processing. An intensive research over all these goals lasts for almost four decades. No known proposal under the traditional paradigm allowed for general purpose encrypted cloud SQL DBMS yet. Only somewhat homomorphic schemes, e.g., the additively homomorphic one of Paillier, appeared fast enough for specific applications. See [JLS5] for more on the related work.

Our paradigm led to a new homomorphic encryption schema called THE-scheme [JLS5]. THE-scheme requires the plaintext basically limited to so-called *core* domain [0.01, 0.02…1M] or similar. The domain is money-oriented. It seems sufficient for many, perhaps most, cloud (outsourced) DBs. THE-scheme allows for standard SQL expressions with any number of the operators '+', '-', '*', '/' and '^' and Count, Sum, Avg, Var and Std aggregate functions. The user may add functions at the expense of storage overhead. All calculations occur at the cloud, unless an overflow or an underflow of the core domain happened. There is no known practical functionally equivalent homomorphic scheme for the core domain.

THE-scheme uses sensitive metadata, called *client secret*. The client sends the secret to the cloud with each query involving arithmetical operators other than additive. The cloud DBMS uses the secret to instantiate some run-time variables with data of interest to the query, selected from their encrypted cloud storage representation. The run-time values can be seen as partly decrypted, called de-obfuscated using the secret, in the terminology of THE scheme. The DBMS discards all the run-time values at the end of the query processing at latest.

Below, we explore the paradigm further. As for THE-scheme, we consider that a query may carry sensitive metadata. These are now however basically the encryption/decryption key(s). A Select query may use an on-the-fly (full) decryption of a DB ciphertext into a run-time plaintext at the cloud. For Update queries, the on-the-fly encryption there may in turn produce the ciphertext going back to the DB. At the query processing termination at latest, the cloud DBMS clears every metadata and plaintext. Under these principles, any relational operation possible over a plaintext, functionally applies to the ciphertext. The DB behaves functionally as if a fully homomorphic encryption was used.

We now call *trusted* the (cloud) DBMS designed as outlined. We believe trusted DBMSs a promising goal for cloud DBMS research. Below, we start with the proposal of the reference architecture. We suppose it software-only, i.e., without any specific (trusted) hardware add-on box. The idea is to apply to a typical cloud node. Next, we suppose that any DB managed under a trusted DBMS is client-side AES-256 encrypted. We call it (cloud) AES DB, managed by a (trusted) AES DBMS. We define the deterministic and a probabilistic AES-based encryption schemes for an AES DB. The former may be faster for an SPJ query, hence preferable for OLTP. It may also or alternatively offer a lower storage overhead. The well-known caveat of any deterministic scheme is potential vulnerability to frequency analysis. A probabilistic scheme is free from the caveat. It is usually nevertheless rather for OLAP, as a deterministic encryption usually does not speed-up aggregations. We define a simple probabilistic version of AES. This one encrypts the blocks where at most half are the plaintext application data. The other part is a random padding (salt). We finally overview the rules for a query execution plan for an AES DB.

We then address the performance analysis. We focus on the processing (time) and storage overhead of queries to an AES DB with respect to the plaintext DB, i.e., with the same plaintext client data. We show that on modern multi-core processors, the popular Intel I5 especially, both processing and storage overhead may be negligible for a deterministic encryption. For the probabilistic one, it should still be so for the processing overhead of aggregations. In contrast the storage overhead at least doubles. We base our conclusions on recent AES benchmarks, [S5] especially. Some show nevertheless the need for experiments targeting specifically a cloud SQL DBMS. These remain a future work.

The end result at present is that the query evaluation over our AES ciphertexts may be usually about as fast as for the plaintext SQL queries. For additions especially, experiments have shown also that the result is almost two orders of magnitude faster than the additively homomorphic Paillier encryption. That one appears the most popular homomorphic scheme at present. Our results show thus, for instance, that 100K SQL additions in Select Sum(x) From… query, may take 14ms over on-the-fly decrypted AES ciphertext. The decryption overhead may be as low as 0.2ms. For Paillier scheme, the result was almost 1.2s, i.e., over eighty times slower. Besides, Paillier scheme needed 50ms per integer sent to the client for the final decryption. All above discussed facts considered, we feel our proposal of potentially major practical importance. Especially, since both heavy weights of cloud DB industry that are Google Cloud SQL and MS Azure SQL offer already the server-side AES 256 transparent encryption. It is however well-known that this one may be unsecure with respect to the cloud service insiders, as the key is always at the cloud. Azure proposes therefore also the client-side encryption. However, this one requires the cumbersome download/upload to/from the client of entire tables to encrypt or even of the entire DB, [S5]. There are also other limitations, e.g., no equijoins on probabilistically encrypted columns [HS5].

Below, next section presents AES DBMS. We first discuss the reference architecture of a trusted DBMS. Then, we show our use of AES for an AES DB. We also discuss the generation of an execution plan for a query to an AES DB. Section 3 analyzes the processing and storage overhead. We conclude and discuss further work in Section 4. We stress that to start practicing with an AES DB appears surprisingly easy. An existing cloud SQL DBMS with UDF (user defined function) support should do.

## 2. The AES DBMS

### 2.1. Reference Architecture of a Trusted Cloud DBMS

Figure 1 presents our reference architecture. We consider it for any cloud DBMS designed under our paradigm, regardless of the cloud DB encryption scheme. In other words, AES DB is a specific case of the architecture, intended as the general one for a trusted cloud DBMS. The DB administrator (DBA) at a client (site) initiates the cloud DB through some kind of upload not discussed here. The DB is encrypted since presumably in insecure storage. Clients manipulate the DB through SQL queries. An SPJ query may end up manipulating a deterministically encrypted ciphertext only, as it will appear. Most queries are nevertheless expected to need plaintexts decrypted at the cloud. An update may require further re-encryption. The cloud DBMS performs both operations on-the-fly, while reading/writing the DB to/from run-time plaintexts. The query needing plaintexts at the cloud brings metadata including the key(s) and info about whether the encryption of a column referred to is deterministic or probabilistic. All the data exchanged are encrypted for the transport using some usual tool (SSL, RSA, Diffie-Hellman…) The DBMS instantiates run-time variables with the metadata brought-in. It deletes any sensitive run-time content, i.e., the metadata and any retrieved/calculated plaintext data, at most by the query processing end.

The cloud processes the queries using some *core* DBMS. This is some full-fledged "classical" plaintext scalar functions for the plaintext DBMS. Our *trusted* cloud DBMS is the core DB reinforced whenever needed through security oriented software (only) re-engineering. The rationale is to allow a trusted DBMS to run on any typical cloud node hardware, i.e., without necessitating some dedicated hardware add-on (box, circuit, trusted computing module (TCM)… [TCG]). The re-engineering is expected to create a protection one may think of as a vault, armor, shield, firewall…. Whatever the name is, the component should protect the core DBMS from exploits disclosing run-time values.

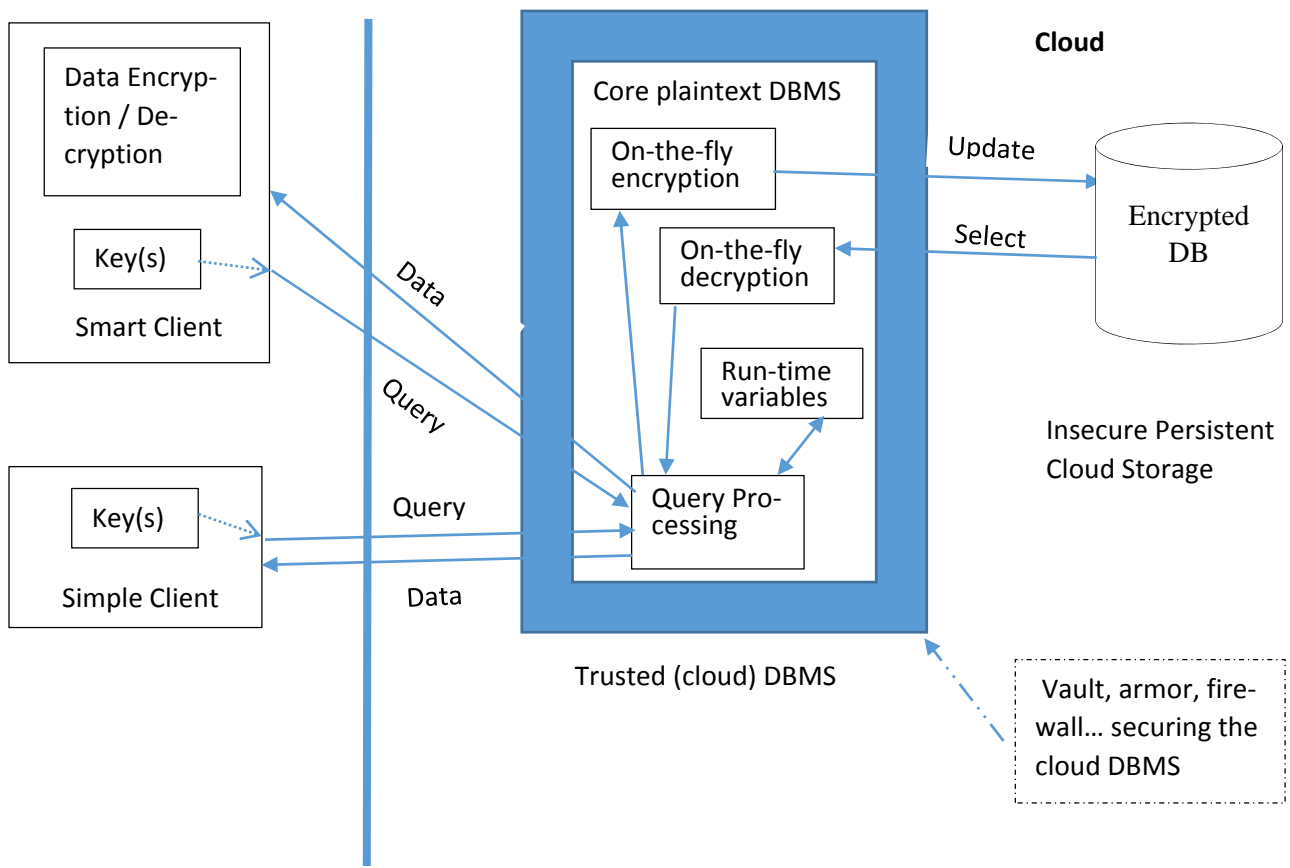

Fig. 1 Reference Architecture of Trusted Cloud DBMS

We believe trusted DBMSs a promising goal for cloud DBMS research. As mentioned, one technical basis for the trust may be a *moving target defense*. Such defenses are extensively discussed in [J&al]. They dynamically and secretly shuffle the storage representation and the location of the program instructions and run-time variables. The concept seemingly traces back to a decade old proposals for secure VMs, [H&al04]. Alternatively, many, perhaps most, users, may possibly turn out happy with a trusted DBMS running on a cloud node simply as usual today. The protection provided in this way may furthermore gradually increase through the software re-engineering taking advantage of the new versions of the popular hardware. E.g., Intel announced in 2013 its Software Guard Extensions (SGX) for the likely most popular world-wide I5 and I7 processors. One expects the new generation of those marketed in late 2015. The processor guards then some RAM areas called *enclaves*. A software outside an enclave cannot read or write the enclave's content regardless of its current privilege level and CPU mode (ring3/user-mode, ring0/kernel-mode, SMM, VMM, or another enclave).  An enclave may be (reasonably) trusted*.* A trusted DBMS security should be able to profit from all this.

While the above and other techniques and developments are promising, a trusted cloud DBMS should not be reasonably expected secure once and forever. A DBMS trusted by some, may not be by others. Stronger protection will likely not come without cost. As for universally trusted OSs, browsers, VMs… the trusted DBMS technology will finally rather face an endless race between proposals, exploits, patches to the exploits and so on.

The trusted DBMS may send the selected data as ciphertext or plaintext or mix both. The ciphertext requires the post-processing decryption at the client. Unlike under the traditional paradigm, a client of a trusted DBMS may alternatively request the entire decryption at the cloud. Avoiding securely the decryption burden may clearly make many clients happier. In practice, even the popular browsers should thus be apt to serve as clients of a trusted DBMS. All this appears an advantage of the new paradigm. The traditional one obviously excludes such clients.

As the result, two kinds of clients and queries appear, Fig. 1. A *smart* client of a trusted DBMS is basically a client DBMS able to locally encrypt/decrypt.  A *simple*, (thin, dumb…) client has not these capabilities. A smart client may send a *ciphertext* query. Such a query uses explicitly the encryption function on ciphertext constants or columns. It may carry ciphertext constants or bring ciphertext to the client for final decryption. A simple client emits *plaintext* queries only. Those cannot bring a ciphertext. The cloud DBMS interprets every constant of a plaintext query as a plaintext and every column name as referring to a plaintext. A plaintext query must include the metadata with the key(s) (wait till Conclusion section for the discussion of somehow alternative convention). Every encryption/decryption is done on-the-fly at the cloud. A ciphertext query may avoid the metadata, as it will appear. It might not need the decryption/encryption on the cloud.  This is a potential security advantage.  Coming however, we recall, at the price of burden at the client. A smart client may of course formulate any plaintext query as well, sending then the key(s). When there are several keys, e.g., one per column, a smart client may also send a ciphertext query with only some but not all the keys required by a plaintext one.

Ex. 1. (a) The following SPJ query is a plaintext one, with the SQL ';' followed by the metadata with the key 'ABCD':
 (1)  Select C.Name, A.Transactions from Customer C, Account A where C.Id = '123' and C.Id = A.Id ; 'ABCD'

To evaluate the query, the cloud DBMS may perform on-the-fly encryption of the selection constant '123'. Or, it may on-the-fly decrypt the visited Id ciphertexts. The choice depends on whether Id encryption is deterministic or probabilistic, as it will appear.  In every case, the DBMS decrypts the final projection (C.Name, A.Transactions) by default.

(b) We denote the encryption function as AESE. The next query could be a ciphertext one of some smart client:

(2)  Select Into Client.CacheDB.CipherTbl AESE (C.Name), AESE (A.Transactions) from Customer C, Account A where AESE (C.Id) = AESE ('123') and AESE (C.Id) = AESE (A.Id) ;

Unlike query (1), query (2) does not carry the key. As said, it is potentially a security advantage over query (1). In contrast, as we detail below, query (2) is valid only if the encryption of Id column is deterministic. Also, it is up to the client to encrypt '123'. If the clause was simply C.Id = '123', it would mean for the cloud necessarily the on-the-fly decryption of Id to test the match of '123'. The query would be rejected as not carrying the key.  As is, the query brings the ciphertext. The result goes to CipherTbl table. The query dynamically creates it in some cache DB for the cloud DB at the client, named CacheDB. The client must decrypt the result for the user (application). E.g. through the following local query, calling the decryption function, say AESD:

Select AESD (C.Name), AESD (A.Transactions) From CipherTbl ;

## 2.2. Deterministic Encryption for an AES DB

From now on, we consider specifically the on-the-fly encryption/decryption through AES, as in Ex. 2. In other words, we restrict our focus to AES DBs only. As usual, we distinguish between OLTP and OLAP queries. It is well-known the former perform best under a deterministic encryption. We recall that every encryption of a plaintext provides then always the same ciphertext. Provided the client-side decryption and the individual encryption of each column plaintext value, i.e., no grouping of column plaintext values into a single ciphertext (see the future work in Conclusion section), the cloud DBMS can evaluate the selections and equijoins of a typical SPJ query over the ciphertext. This could be the case of query (2) provided Name, and Id column deterministically encrypted. Transactions column may or may not be deterministic. The client must be a smart one.

We recall that AES is a symmetric basically deterministic scheme that encrypts plaintexts into 16B blocks. Modern processors shifted to 64b arithmetic on typical workstations. We thus consider basically that a numerical value is up to 8B long. A string may have any length. If the ciphertext-based evaluation is allowed for SPJ-queries, then the client encrypts each plaintext as a distinct ciphertext. If the plaintext is numerical the client trivially pads it to the block length before encryption. For instance, the padding adds the leading zeros. Likewise, if a string length is not a 16B multiple, the client pads the partly filled block, e.g., with spaces. The padding, say $p$, has simply to be such that for any plaintext $x, y$, if $x = y$ then AESE ($p, x$) = AESE ($p, y$), i.e., both ciphertexts are equal. Each ciphertext is then clearly a deterministic one. The cloud DBMS knows the padding by default or gets info in the metadata.

Having a single numerical plaintext per block should usually create storage overhead with respect to the plaintext DB. To reduce this one, the client may group column plaintext values into a single ciphertext to fill up each bloc, as already mentioned. Either row-wise or column-wise, depending on DBMS implementation. In every case, a ciphertext only based query evaluation referring to a grouped column is not possible (for AES DB, but perhaps not for a cloud DB encrypted differently; see the discussion of Paillier scheme later on). The query must bring the key with.

The latest version of the AES standard is AES 256, using the 256b key over the 128b blocs. We focus naturally on this version. Available benchmarks do so as well.

## 2.3. Probabilistic Encryption for an AES DB

We recall that a deterministic encryption may be vulnerable to frequency analysis. It is well-known that it is secure only on a low-entropy domain, i.e., with many about uniformly distributed domain values. Thus may be OK for a presumably random Id above, or SS# or TaxPayer# etc. It may not be so for above Transactions or ZIP code or Salary etc. Generally, it may be not OK for any column with skewed domain value distribution. A client may make such columns low-entropy anyhow through additional pre-processing, e.g., adding *decoys*. This may be a burden for some.

A probabilistic encryption encrypts a plaintext into a large number of seemingly random ciphertexts. AES is basically deterministic. We turn it to a probabilistic encryption for a numerical plaintext, by simply filling each bloc to encrypt with the plaintext up to 8B large. Say putting it to the right half. The left half of the block contains client chosen random padding. For a string, the client does it for each block that otherwise would be filled up. For the last, eventually incomplete, block, a fix padding may eventually fill up the half with the plaintext before the encryption.

To one plaintext in any block, may correspond equally-likely at least $2^{64}$ AES ciphertexts. The seemingly random ciphertext produced in this way by a probabilistic encryption, in practice precludes any frequency analysis. In turn, it basically impairs SPJ queries. E.g., consider the following query requesting the selection of all other clients with the ZIP code of a given one: (3) Select X.Id, X.Name From Customer as X, Customer as Y Where X.Zip = Y.Zip and X.Id = '123' and Y.Id <> '123'. If ZIP is probabilistically encrypted, very likely, there is no other client with a probabilistic ciphertext of ZIP of '123'. Actually, even if the plaintext ZIP of '123', say 60520, happened to be cached at the client, any re-execution of AESE (60520) would lead to a different ciphertext. Bottom line is that there is no practical way to evaluate a join directly over a probabilistic ciphertext, regardless of the encryption algorithm used. A trusted cloud DBMS would be unable to evaluate a plaintext query (3) or any other with the join over probabilistically encrypted columns, if the key was not coming with. While it would be if ZIP encryption was deterministic and a smart client rewrote the query as a ciphertext one, similarly to query (2).

A query to an AES DB referring to a column with a probabilistic ciphertext, may be a plaintext one. E.g., like query (3) but with the key in the metadata. The client does not need then to know what the encryption of a column is really. Smart client may alternatively issue a ciphertext query, provided it includes the key as well. The DBMS is obviously functionally able to evaluate any join or selection through their on-the-fly decryption entirely at the cloud.

Notice here that there is another widely known functional possibility to evaluate a query over a probabilistic encryption. The cloud DBMS may send a possibly minimal collection of encrypted columns to the client for decryption and plaintext post-processing. The DBMS does not need the key what makes this approach the only functional possibility for a cloud DBMS under the traditional paradigm. The size of the collection makes however this strategy usually impractical, as also well-known. We do not consider this approach for a trusted DBMS further, although it may be perhaps an option sometimes.

### 2.4. Query Execution for an AES DB
With respect to the execution plan, any trusted DBMS, the AES DBMS in particular, basically blends the on-the-fly decryption into the execution plan optimal for the DB as if it was the plaintext one. In other words, it generates some plan as if the DB contained the plaintext. Then, it decrypts on-the-fly whenever needed every ciphertext selected during the query execution. In this way, the DBMS is able to execute for any encrypted DB any query valid for the plaintext one.

The deterministic ciphertext may or may not need the decryption, as we have seen. The probabilistic one always does. Deterministic ciphertexts can have indexes. The plan may depend also on whether the ciphertext groups the individual plaintext column values, as already hinted. The grouping can be then row-wise or column-wise. This depends on whether the physical tables (files) have the records with rows or have a columnar structure. We recall that the former, e.g., of MySQL is OLTP oriented, while the latter aims at OLAP queries, e.g., for Vertica or Monet DBMSs. Details of query optimization remain a future work.

Ex. 3 Suppose that (a) the client sends query (3) as a plaintext one, (2) the physical Customer table has the row-oriented structure, (3) columns are not indexed, (4) every column plaintext value there is individually encrypted and (5) both Id and ZIP columns ciphertext are probabilistically encrypted. A plan A may be to start with the scan of Customer for the selection of the row where Id = '123'. This involves on-the-fly decryption of Id ciphertext to locate the row with '123'. Suppose that it happens and that Id is the key of Customer, hence there can be only one such row. The plan may then decrypt ZIP there and store in some run-time variable, say V. Then may continue with the 2$^{nd}$ scan through all the rows of Customers but the one found, decrypting on-the-fly each visited Id and ZIP and attempting the match with V as specified by the query. If the match succeeds, then the plan decrypts NAME and stored the row (Id, Name) in a run-time table T. Once the scan ends, the DBMS sends T to the client and clears all the run-time variables.

A plan B may involve a single scan of Customer only. In each visited, DBMS decrypts on-the-fly Id, Name and ZIP. If 'Id' <>, then it puts the plaintext aside into run-time table T1 (Id, Name, Zip). Once it finds the row with '123', it puts its ZIP aside into some variable V. Then continues till the end the scan of the Customer, but creates the run-time table T2 (Id, Name) from the Customer rows that have ZIP = V only. Once done with the scan, it sends out T3 = T1 (Id, Name) Union T2. Obviously, whether plan A or plan B turns out the better one depends on the data and memory cost. Plan B needs more but saves the time for one scan. Both plans would be slightly different if one ciphertext groups all the plaintext in a row. Other plans may be preferable, e.g., if Id is deterministic hence may not always need the decryption. Or if it is also indexed etc.

Ex. 4. Consider the query:

(4)  Select Sum (Transactions), INT (Var (Transactions)) From Account;

Suppose the client sends it as a plaintext one with some metadata with the key and that Account (cipher)table is row-structured. The execution plan may scan Account decrypting-on-the fly each Transaction ciphertext. It then may as usual calculate both aggregates over the run-time plaintext. Sending finally the plaintext result to the client and clearing all the run-time variables. Except, for the decryption, this is simply the basic usual execution plan for query (4) addressing a plaintext row-structured Account table.

It is worth noting that to execute query (4) at the cloud under the traditional paradigm, the encryption of Transactions must be fully homomorphic. There is no such scheme providing even remotely practical response time as yet. THE schema should be often practical for query (4), provided that the plaintext are limited to the domain dealt with. Otherwise, the only plan can be to send entire Account or at least its Transactions column to the client for the post-processing. This plan is however known as impractical for tables beyond the toy ones at present. Notice that it precludes also the simple clients.

Last, but not least, we recall that if query (4) was limited to SUM aggregate, it could however execute at the cloud under the traditional paradigm with often acceptable processing time. This, provided the Paillier's additively homomorphic probabilistic encryption on Transactions, [S&al4]. This one may work over grouped plaintext values as well. We'll come back to the Paillier's scheme later on. We'll show that even in this limited case, our scheme should be orders of magnitude faster.

## 3.   Performance Analysis

### 3.1.   Processing Overhead

What matters most for our proposal is the overhead of on-the-fly decryption and encryption induced by a query evaluation at the cloud over the ciphertext in AES DB. This, with respect to the query evaluation over the plaintext. There are several recent benchmarks of AES: [S5], [G1], [I2]… These consider the popular multi-core processors. Most of them naturally considers the ciphertext in RAM cache or disk. The encryption/decryption result can be measured as sent out (or simply dropped) or with every ciphertext/plaintext written back to RAM. The former measure is the basic one for Select queries. The latter one adds up for systematic Update queries. For instance adding 10% to every price in some table. The main measure is the number of encrypted/decrypted bytes per second (MBs). The decryption can be little faster than encryption.

The encryption can be entirely in software. Two popular public-domain algorithms are Truecrypt and Twofish. The former uses the Rijndael's algorithm that won NIST competition. The latter was a competitor as well, but appeared slower, for 64b processors especially, [S&al]. Within Intel I5 processors family, several CPUs have instructions for AES encryption/decryption hardware acceleration. These are so-called AES-NI instructions. Some Xeon CPUs also do, e.g., Xeon X5690. Pricing with or without NI is in practice the same. Truecrypt 7.0a takes advantage of AES-NI. Twofish does not. The benchmarks show that AES-NI effectively speeds up the processing. Results vary among benchmarks.

For our purpose, we concentrate on I5, as the most used. According to [G2], the bulk raw (straight) encryption using the Truecript 70.a without RAM re-writing provides the impressive 1900 MBs encryption/decryption rate. Twofish leads to 273 MBs "only". More recent results in [S5] for a wide range of CPUs, report for I5 661 CPU specifically, an even more impressive 4133 MBs rate. Presumably, with Truecript 70.a as well. Results for other CPUs vary, the slowest being 317 MBs and the average being 1.9 GBs. For the deterministic encryption this leads up to 516,5M for AES-NI and to 34M for Twofish pf plaintexts/ciphertexts processed per second. To decrypt 100K values, e.g., for sum SUM function, may take thus as little as 0.2 ms with Truecript 70.a (and 3ms with Twofish). For our probabilistic encryption, the timing multiplies by two.

The processing naturally slows down when every decrypted/encrypted value is written back to RAM. Only [G1] reports the related experiment, using Truecript 70.a. It performed at 763 MBs. However, the plaintext writing rate was then limited to 880 MBs. Encrypting led thus to 13% overhead only. Per value rate is about 100 - 50Ms and 100K value decryption takes 1-2ms for our encryptions. How the RAM writing impacts an SQL query depends obviously on the aggregates and clauses (GROUP BY, ORDER BY TOP…). Nevertheless, Select queries serve generally to produce few values only. An aggregate is expected to read perhaps very many tuples, but to produce from relatively a few only. The writing timing of these results should therefore very little impact of the read-only results above. It is not the same for a large update. We come back to the issue below in SQL specific analysis.

The bulk transfer rate from hard or SS (flash) disk is disk technology dependent. They appear to be at most 150MBs in practice (SATA-3 interface). The random access times are well-known, i.e., about 10ms in practice for a hard disk and 1ms for an SSD. The AES overhead appears negligible, allowing for the real-time processing (Aegis Padlock disks).

The results for the decryption/encryption of selected values or of small groups of those are slower than for bulks. The reason is so-called *key set-up* time. Experiments show nevertheless that the key set-up may cost for the Rijndael's algorithm as little as 15% slow-down [S&al]. An SQL query is typically expected to do a bulk search. We thus neglect this (small anyway) specificity in what follows.

Finally, the EAS algorithms above discussed appear programmed in assembly language. Use of a higher-level compiler, e.g., Java, may have a severe impact. For Oracle JDK 1.7, Intel reports thus at best 80 MBs rate, for AES-NI, [I5]. This is 10M values per second for us, "only". The overhead goes up to 10ms per 100K decryptions. We do not analyze this result further. Use of best optimized implementation for a cloud DBMS seems natural. The subject requires nevertheless a specific study.

### 3.2. Storage Overhead

Our deterministic AES scheme may have no storage overhead. As discussed, it is however unlikely in practice that any AES DB loads all its blocks exactly. Also, if plaintext values of a column are individually encrypted, to allow for SPJ queries over the ciphertext, each numerical value "expands" to 16B from 4-8B usually. Practical overhead should depend on the DB scheme and on the design choices for the encryption we hint to in Conclusion section. It remains a future study. Our probabilistic encryption carries at best the 100% overhead, i.e., doubles the plaintext storage. This is nevertheless what, e.g., a probabilistic homomorphic scheme may need at least as well, e.g., the Paillier's scheme. So our scheme is not worse on storage requirements, after all.

### 3.3. Query Processing

The basic measure of this one is the overhead of on-the-fly decryption on the otherwise plaintext execution plan for the same SQL query. We have seen several example plans. The overhead may clearly depend on a plan. Globally however, we have seen that the decryption may deal even with GBytes of data per second. Also, the study of the read/write speeds for a ciphertext and a plaintext above has shown only 13% overhead. All this suggests that even for a RAM DB, the overhead of the on-the-fly decryption on the execution could be usually limited to a dozen of percent or so as well. It should become negligible if the DB is solid state or hard disc resident. More precise evaluation remains a future goal, except for what immediately follows.

Another measure can be the query execution speed of a query with respect to the same query to a homomorphic encryption. Provided of course that the query is then executable. With respect to this approach, [S&al3] studies the SQL query processing timing for SUM aggregation for both plaintext and Paillier scheme. This scheme is considered the most efficient for an additively homomorphic probabilistic encryption at present. See [S&al4] and various references in [JLS5]. The Select SUM(x)… query adding 100K plaintext values at the cloud server needed 14ms. Our scheme could add up at best 0.2ms as above discussed. The overhead could thus be as low as 1.5 %. All other benchmarks we cited would cost only a few milliseconds at most.

For the same query, the reported scan timing for the Paillier ciphertexts is 1.16s. This is almost 83 times the plaintext timing. About the same ratio, namely almost 82 times for the deterministic choice and 81 for our probabilistic one, may thus apply to an AES DB. The decryption on the client requires further 50ms per integer for Paillier. This is not a problem for a few value. But definitively could be for more. For instance, a Group By aggregating even a thousand values to one for our 100K tuples, would need 50s for the decryption. Such response times are impractical for many applications. For an AES DB, in all the discussed cases, the client decryption time could be a fraction of a millisecond as well. We recall that the client of an AES DB decrypts only when the query sends back a ciphertext and that the client can always request the entire decryption at the cloud.

### 5. Conclusion

On-the-fly decryption/encryption by a trusted cloud DBMS, appears the first generally practical architecture for a relational cloud DB. It is the only to offer in practice at present all the functional capabilities of a plaintext relational DBMS. It is also the only to allow for simple clients. The on-the-fly decryption/encryption run-time overhead should be negligible for an AES DB, i.e., using our deterministic or probabilistic AES encryption. It should be also only a small fraction of any known homomorphic encryption scheme. Furthermore, the functional and processing capabilities of known homomorphic scheme perhaps sufficient for selected applications, are all also largely limited with respect to our scheme. Our proposal appears to reach the "Holy Grail" of intensive research for almost four decades. We feel it worth further investigation.

That study should at first concern the design of an AES DB. There are specific optimization issues. We may call one, already signaled, the *plaintext grouping*. The issue is to form for a give DB the optimal single AES ciphertext. First question is: should it be row or column-wise? If a usual query is the OLTP oriented one; like Select * From Where…, the row-wise grouping seems most efficient. Perhaps, however, there should be no grouping at all, i.e., singe plaintext per ciphertext? We recall that this choice lets the DBMS to evaluate SPJ queries over ciphertexts, provided the client-side decryption. In turn, the storage overhead is adversely affected. Finally, the grouping could be physically oriented, e.g., one ciphertext per file page (block, bucket…).

A related issue is that of the *key granularity*: one key per whole DB, per table or perhaps even per column? A client may prefer the latter option for security, [D&al05], but it is also obviously most cumbersome. Also, the metadata field carrying the key(s) we considered for a query to an AES DB, while easy to use for a single key per table, is no more so if the granularity is a column. Besides, it is not known to a plaintext SQL DBMS at present. This could hamper the implementation of an AES DBMS. An alternative convention for a query to an AES DB could be to rather make each key an argument of the decryption function AESD. E.g, the plaintext query (1) could carry three different column keys, becoming for deterministically encrypted Id perhaps: Select AESD (C.Name, ABC), AESD (A.Transactions, BCD) from Customer C, Account A where AESD (C.Id, CDE) = '123 and C.Id = A.Id;. Future work should determine best choices.

Next, the DBA should have rules to weigh between the deterministic and probabilistic encryption. As we discussed, the choice may influence the security and storage overhead, but also the query processing overhead and even query writing or rewriting. For instance, in the above query, if Id encryption is probabilistic, the above selection clause clearly can be only as it is. For the deterministic encryption, the clause stays valid as well. However, the DBMS should rather rewrite it as C.Id = AESE ('123', CDE). The encryption/decryption processing overhead should become many times smaller usually, i.e., *N* times smaller for *N* Id ciphertext values visited. Next, the join clause would not work as is at the cloud, referring by default to ciphertext, we recall. Either the user should write it as or the DBMS should rewrite to: AESD (C.Id, CDE) = AESD (A.Id, CDE). The rewriting is possible here since the Id key is already elsewhere in the query. It would be impossible otherwise.

Another issue is that of the best file structure for an encrypted table, column or index. The key-order does not seem to matter for an AES encrypted file. Hash is perhaps preferable over otherwise ubiquitous B-tree. A plaintext SQL DBMS may offer a choice, e.g., see Postgres for options of CREATE statements. All these issues affect the processing speed so should be studied.

Next, there are query execution plan generation issues. Any plan has to take to the account the encryption type on a column, whether a query or its part can be evaluated directly over encrypted values etc. We have seen a few plans, but, as already said, rules for best plans for an AES DB remain to be studied. The outcome may affect the design of AESE and AESD functions. Finally, one should experiment with the trusted AES cloud DBMS actual processing speed for an SQL oriented benchmark, e.g., a TPC benchmark. In the secure software engineering arena, one should continue making the trusted DBMSs more and more secure. As pointed out, no trusted system can be expected secure once and forever.

All these ambitious goals spelled out, if one trusts an existing plaintext cloud SQL DBMS, making it a basic AES DBMS may appear astonishingly easy. The scalar functions MySQL AES_ENCRYPT() and AES_DECRYPT() in free version of MySQL may serve as our AESE (c, k) and AESD (c, k) functions for the columnar granularity. The RAND function may help with the probabilistic encryption. More generally, our goal seems also easy if an existing plaintext cloud SQL DBMS supports user defined functions (UDFs), (unlike, e.g., Google Cloud version of MySQL at present). SQL Server seems a good candidate as MSDN already offers the AES 256 encryption/decryption routines. In each case, a browser suffices to run plaintext queries as a simple client. It is likely the way to start practicing with our proposal.